# The Guidebook, the Friend, and the Room: Visitor Experience in a Historic House

**Allison Woodruff, Paul M. Aoki, Amy Hurst, and Margaret H. Szymanski**
Xerox Palo Alto Research Center
3333 Coyote Hill Road
Palo Alto, CA 94304 USA

**ABSTRACT**
In this paper, we describe an electronic guidebook prototype and report on a study of its use in a historic house. Supported by mechanisms in the guidebook, visitors constructed experiences that had a high degree of interaction with three entities: the guidebook, their companions, and the house and its contents. For example, we found that most visitors played audio descriptions through speakers (rather than using headphones or reading textual descriptions) to facilitate communication with their companions.

**INTRODUCTION**
Visitors to cultural heritage locations have multiple goals. Often they want to get information about the objects they see. However, sharing the experience with their companions is often a higher priority than education, particularly for infrequent visitors [2]. Similarly, a "romantic" experience, one in which the visitor "gets the sense of the place," is often prioritized more highly than education by both curators and visitors [1]. Therefore, visitors generally divide their attention between at least three entities when visiting a cultural heritage site: (1) an information source (e.g., a guidebook); (2) their companions; and (3) the location itself.

An electronic guidebook must be designed to fit appropriately into the visitor's desired experience. More specifically, designers should be aware that visitors will make behavioral choices with respect to all three attentional entities, and that the guidebook can either support or hinder these efforts. Previous work on context-aware applications has focused on methods for presenting information; little work has been done on questions of attentional balance.

In this paper, we report on a study of electronic guidebook usage. We first constructed a prototype, designing it to provide a range of options for information presentation and sharing. We then observed fourteen visitors using the guidebook in a self-guided tour of a historic house. Our analysis here focuses on how the device helped visitors balance the competing demands of multiple attentional entities.

**PROTOTYPE**
The electronic guidebook application runs on a Casio Cassiopeia E-105 color personal digital assistant (PDA). It presents the visitor with one of a collection of imagemaps. Each imagemap is a photograph taken facing one wall of a room in a historic house. The visitor changes the viewing perspective (i.e., displays a different imagemap) by pressing a button on the device. When the visitor taps on an imagemap target, the guidebook displays a brief text description or plays an audio clip of the same information. The visual navigation design is motivated by the principles described in [1]. Location-aware systems are not feasible in historic houses for a number of reasons, e.g., barriers often prevent visitors from approaching objects. Usability testing of the prototype by thirteen users confirmed that visual selection is a viable alternative that allows visitors to quickly and easily select objects that interest them.

The prototype gives the visitor several choices with regard to information presentation. The choice of audio or textual presentation can be changed at any time. Audio clips can be played at a low volume through speakers on the device or through headphones. These choices support a range of privacy and sharing options.

**METHOD**
The study participants were members of the Xerox PARC community, accompanied by friends or relatives with whom they would normally attend a museum. For example, a grandmother attended with her 7-year-old granddaughter and a husband attended with his wife. The visitors comprised a total of seven couples and ranged in age from 7 to over 60 years of age (two of the couples included a child as a visitor; all other visitors were adults). Many of the visitors were non-technical and/or had not previously used a PDA.

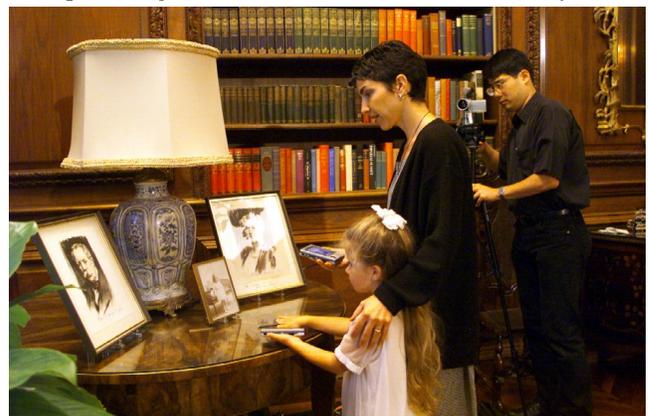

Figure 1: Visitors using electronic guidebooks at Filoli.

Participants were observed during a private visit to Filoli, a Georgian Revival house in Woodside, California. The visitors went through the first several rooms of the house with a paper guidebook, accompanied by a docent who was available to answer questions.

The visitors used the electronic guidebook in the next two rooms of the house. The visitors received brief instructions in the use of the guidebook. They were then asked whether they would each like their own guidebook or if they would prefer to share. They were also offered headphones. They were told that they could change their decisions at any time. The visitors then spent approximately 20-30 minutes (without a time limit) in the two rooms, referencing the electronic guidebooks as desired. The visitors' comments were recorded using wireless microphones, the visitors were videotaped by a camera in a corner of each room, and the visitors' actions in the electronic guidebook were logged for future reference. After they were finished in the two rooms, a semi-structured interview was conducted.

**FINDINGS**
Visitor response to the electronic guidebook was extremely enthusiastic. The electronic guidebook had a substantive impact on visitors' interactions with each other and with the rooms and their contents. Further, visitors chose guidebook options that facilitated these interactions.

*Common models of guidebook use:* Each member of the five adult-only couples chose to use their own device. Four of these five couples predominantly used audio played through the speakers of the device. In the other couple, one visitor played audio through the speakers of his device, and his companion listened to audio through headphones.

One couple consisting of a child and an adult shared a single device during the entire visit, with the child operating the device (playing audio descriptions through the speakers) and the adult looking on and making suggestions. The other couple with a child and an adult began by using separate devices, but gradually evolved to the same shared model as the other child/adult couple.

Although audio was the primary medium, some visitors reported switching to text for specialized purposes, e.g., to learn how a word was spelled. Some visitors felt that a combined text/audio mode, in which both text and audio were presented simultaneously, would be useful.

*Visitor-visitor interaction:* The use of audio clips played through the speakers facilitated visitor-visitor interaction. This is in contrast to the use of audio in headphones, which visitors often report isolates them from their companions.

Visitors would frequently stand right next to each other playing different audio clips through the speakers. Visitors had a high tolerance for these overlapping audio clips, some stating that they "browsed" the audio clips being played by their companions' devices. Also, visitors often played audio clips for each other. One visitor initially repeated information from text descriptions to her companion, but then realized that playing audio clips was a more effective way to get her companion's attention. She switched to audio mode, saying, "[I] have to play the audio to get your attention, huh?" One couple briefly used two devices to simulate a combined text/audio mode, one member instructing the other, "You get the text and I'll get the audio."

The content of the audio clips often served as a springboard for conversation, either about the objects themselves, or about related topics, e.g., a discussion of how the family spent Christmas at Filoli led one visitor to remark on an upcoming Thanksgiving celebration of her own.

*Visitor-room interaction:* We initially wondered whether the device would dominate the visitors' attention. However, visitors spent a great deal of time looking at objects in the room, aided by the fact that the guidebook required little attention to operate and by the use of audio (as opposed to text) descriptions. Many visitors stated that they preferred audio because it allowed them to look at the room at the same time they were getting information. The content of the descriptions influenced visitor behavior. For example, one description mentions detailed carving on a fireplace. When visitors heard this description, they often walked across the large room to inspect the carving.

**CONCLUSIONS AND FUTURE WORK**
Most visitors took advantage of mechanisms in the guidebook to incorporate the guidebook, their companion, and the room in their experience. To achieve balance among these competing entities, they made a number of complex tradeoffs. For example, most visitors chose to use their own devices rather than share, but they simultaneously created shared audio environments that allowed them to participate with their companions.

Future work includes further analysis of the data, e.g., we are currently conducting a detailed conversation analysis. We are also investigating new guidebook designs, e.g., designs that provide shared audio spaces using headphones that do not block the ear channel.

**ACKNOWLEDGMENTS**
We are deeply indebted to Tom Rogers and Anne Taylor of Filoli for their generous assistance with this project. We also thank Tom for his perceptive comments on the design of the prototype, Maribeth Back for assistance in recording the audio clips, and Bob Moore for his helpful insights.